\documentclass[aoas,nameyear,rotating,dvips]{arximspdf}
\usepackage{textcomp}

\doi{10.1214/10-AOAS343}
\volume{4}
\issue{4}
\pubyear{2010}
\firstpage{1871}
\lastpage{1891}

\makeatletter

\newcommand{\iint}{\int\!\!\!\int}
\newproclaim{assum}{Assumption}
\newproclaim{rem}{Remark}
\newtheorem{theorem}{Theorem}

\makeatother

\begin{document}
\begin{frontmatter}

\title{Improving PSF calibration in confocal microscopic
imaging---estimating and exploiting bilateral symmetry}
\runtitle{Bilateral symmetry in imaging}
\begin{aug}
\author[A]{\fnms{Nicolai} \snm{Bissantz}\ead[label=e1]{nicolai.bissantz@rub.de}\thanksref{t1}},
\author[B]{\fnms{Hajo} \snm{Holzmann}\corref{}\ead[label=e2]{holzmann@mathematik.uni-marburg.de}\thanksref{t2}}
\and
\author[C]{\fnms{Miros{\l}aw} \snm{Pawlak}\ead[label=e3]{pawlak@ee.umanitoba.ca}}
\runauthor{N. Bissantz, H. Holzmann and M. Pawlak}
\thankstext{t1}{Supported by the BMBF project INVERS and of the DFG SFB
475 and SFB
823.}
\thankstext{t2}{Supported by the DFG Grant HO 3260/3-1, the
Claussen--Simon--Stiftung
and the Landesstiftung Baden-W\"{u}rttemberg, ``Juniorprofessorenprogramm.''}
\affiliation{Bochum University, Marburg University and University of Manitoba}
\address[A]{N. Bissantz\\
Fakult\"{a}t f\"{u}r Mathematik\\
Ruhr-Universit\"{a}t Bochum, NA 3/70\\
Universit\"{a}tsstr. 150\\
D-44780 Bochum\\
Germany\\
\printead{e1}} 
\address[B]{H. Holzmann\\
Fachbereich Mathematik und Informatik\\
Philipps-Universit\"{a}t Marburg\\
Hans-Meerweinstrasse\\
D-35032 Marburg\\
Germany\\
\printead{e2}}
\address[C]{M. Pawlak\\
Department of Electrical and Computer Engineering\\
The University of Manitoba\\
Winnipeg, MB R3T 5V6\\
Canada\\
\printead{e3}}
\end{aug}

\received{\smonth{1} \syear{2009}}
\revised{\smonth{2} \syear{2010}}

%
\begin{abstract}
A method for estimating the axis of reflectional symmetry of an image
$f(x,y)$ on the unit disc $D = \{ (x,y)\dvtx x^2 + y^2 \leq1\}$ is
proposed, given that noisy data of $f(x,y)$ are observed on a discrete
grid of edge width $\Delta$.
Our estimation procedure is based on minimizing over $\beta\in[0, \pi
)$ the $L_2$ distance between empirical versions of $f$ and $\tau
_\beta
f$, the image of $f$ after reflection at the axis along $(\cos\beta,
\sin\beta)$. Here, $f$ and $\tau_\beta f$ are estimated using
truncated radial series of the Zernike type. The inherent symmetry
properties of the Zernike functions result in a particularly simple
estimation procedure for $\beta$. It is shown that the estimate $\hat
\beta$ converges at the parametric rate $\Delta^{-1}$ for images $f$ of
bounded variation. Further, we establish asymptotic normality of~$\hat
\beta$ if $f$ is Lipschitz continuous. %
The method is applied to calibrating the point spread function (PSF)
for the deconvolution of images from confocal microscopy.
For various reasons the PSF characterizing the problem may not be
rotationally invariant but rather only reflection symmetric with
respect to two orthogonal axes.
For an image of a bead acquired by a confocal laser scanning microscope
(Leica TCS), these axes are estimated and corresponding confidence
intervals are constructed. They turn out to be close to the coordinate
axes of the imaging device. As cause for deviation from rotational
invariance, this indicates some slight misalignment of the optical
system or anisotropy of the immersion medium rather than some irregular
shape of the bead. In an extensive simulation study, we show that using
a symmetrized version of the observed PSF significantly improves the
subsequent reconstruction process of the target image.
\end{abstract}
%
%
\begin{keyword}
\kwd{Image analysis}
\kwd{semiparametric estimation}
\kwd{reflection symmetry}
\kwd{two-dimensional functions}
\kwd{Zernike polynomials}
\kwd{confocal microscopy}.
\end{keyword}

\end{frontmatter}

%
%
\section{Introduction}
The fundamental concept of symmetry of physical and biological objects
has been thoroughly studied for a long time; cf., e.g., Conway, Burgiel
and Goodman-Strauss (\citeyear{Conway08}). In particular, symmetry plays an important
role in image analysis and understanding and finds direct applications
in object recognition, robotics, image animation and image compression;
see Liu, Collins and Tsin (\citeyear{Liu04}) for an overview of the subject of
symmetry and related issues. The problem of detecting and measuring
object symmetries has been tackled in the image processing and pattern
analysis literature since the original works of Atallah (\citeyear{Atallah85}) and
Friedberg (\citeyear{Friedberg86}). For a comprehensive review of the literature see
Liu, Collins and Tsin (\citeyear{Liu04}) and Bissantz, Holzmann and Pawlak (\citeyear{Bissantz09}). The role of symmetry
in statistical inference is discussed in Viana (\citeyear{Viana08}).


In this paper we propose an estimation procedure for the angle $\beta$
of the direction $(\cos\beta, \sin\beta)$ of the axis of reflectional
symmetry of an image function $f$ from which discrete, noisy
observations are available. The observations are taken on a grid of
edge width $\Delta$, and the noise is modeled by stochastic errors.
Existing methods either do not allow any noise or treat the effect of
noise only empirically by simulations. Thus, to the best of our
knowledge, our approach is the first which treats reflection symmetry
estimation from a statistical point of view as a semiparametric
estimation problem. Specifically, we show that for image functions
$f$ of bounded variation the estimate $\hat\beta$ converges at a
rate of $\Delta^{-1}$ and, further, for Lipschitz continuous $f$
we have asymptotic normality, which allows us to construct asymptotic
confidence intervals for $\beta$.

The estimation procedure is based on minimizing over $\beta\in[0, \pi
)$ the $L_2$ distance between empirical versions of $f$ and $\tau
_\beta
f$, the image of $f$ after reflection at the axis along $(\cos\beta,
\sin\beta)$. Here, $f$ and $\tau_\beta f$ are estimated using
truncated Zernike function expansions. The inherent symmetry properties
of the Zernike functions yield a particularly simple estimation
procedure for $\beta$. In the recent related papers [cf. Kim and Kim
(\citeyear{Kim99}) and Revaud, Lavoue and Baskurt (\citeyear{Revaud08})], methods for estimating
the rotation angle of an image invariant under a certain rotation,
which also make use of the Zernike moments, have been proposed.
However, the authors do not study any convergence aspects of the
algorithms and confine their discussion to noise-free images.

Our methodology is applied to calibrating the point spread function
(PSF) of a microscope in confocal microscopy.
The PSF describes the blurring effect of the imaging process.
Typical smoothing scales are of order $\approx$100~nm, which often is
of similar order as the size of relevant structures in the target
object. Hence, an exact knowledge of the PSF is essential to properly
adjust (i.e., deconvolve) the observed image to recover the image of
the target object.

A theoretical PSF may be computed from the optical properties of the
microscope, it is rotationally invariant for a rotationally symmetric
optical system.
However, the true (empirical) PSF can deviate substantially from its
theoretical shape, and is no longer rotationally invariant. Therefore,
the PSF is estimated from images of point-like objects with known form.
Since this process involves rather dim images, it is worthwhile to use
additional information on the PSF to improve on its reconstruction.

Often, the empirical PSF is still expected to be reflection symmetric
with respect to two (unknown) orthogonal axes, for example, if the
detector plane is not in perfect agreement with the focal plane of the
microscope; cf. Lehr, Sibarita and Chassery (\citeyear{lehr98}) and Pankajakshan et
al. (\citeyear{Pankajakshan08}).
Therefore, for an image of a bead acquired by a confocal laser scanning
microscope (Leica TCS), in Bissantz, Holzmann and Pawlak (\citeyear{Bissantz09}) we used
hypotheses tests to assess rotational invariance as well as invariance
under a rotation by $\pi$ (which is a consequence of invariance under
reflections by two orthogonal axes) for the empirical PSF. While (for
bead~2) rotational invariance was rejected, invariance under a rotation
by $\pi$ (and hence reflection symmetry) was not rejected at the level
of $5\%$.

Here, we estimate the axes of reflectional symmetry of the PSF and construct
the corresponding confidence intervals.
It turns out that the axes are very close to the coordinate axes of the
imaging device.
This indicates that the reason for the PSF to deviate from rotational invariance
appears to be some (slight) misalignment of the optical system or anisotropy
of the immersion
medium used for object preparation rather than some random deviation from
sphericity of the bead used to image the PSF.

Further, we propose to reduce the noise level in the PSF by a factor of
2 by averaging along the estimated axes.
To investigate the practical merit of this strategy for recovery of a
target image, we use a two-step simulation study. First, the PSF is
estimated by four different methods, then the estimated PSFs are used
for subsequent recovery of the target image, and the accuracy of these
reconstructions are compared.
For the PSF we use a simple nonparametric estimate of the PSF as well
as a symmetrized version, together with correctly specified and
slightly misspecified parametric models. It turns out that while the
correctly specified parametric model performs best for recovering the
target image, symmetrizing the nonparametric estimate greatly improves
its performance, even beyond that of the slightly misspecified
parametric model.

The paper is organized as follows. In Section \ref{sec:zernike} we
introduce the theoretical Zernike moments and give their basic
invariance properties. Further, we discuss how to estimate the moments
from data generated by our observational model. In Section \ref
{sec:reflecest} we propose the estimation procedure for the angle
$\beta
$ of the direction $(\cos\beta, \sin\beta)$ of the axis of
reflectional symmetry of the image function $f$, and discuss its
statistical properties. This includes the issue of uniqueness as well
as consistency, rate of convergence and asymptotic distribution of the
estimate. Section \ref{sec:finitesample} contains simulation studies
concerning the finite sample properties of the estimator. In Section
\ref{sec:simu} we discuss reflection symmetry properties of an observed
PSF of a confocal laser scanning microscope (Leica TCS). Further, in a
simulation we show how incorporating reflection symmetry into a simple
nonparametric estimate of the PSF significantly improves its properties
in the image reconstruction process. Section \ref{sec:conclusions}
gives some concluding remarks, while technical proofs can be found in
the supplementary material in Bissantz, Holzmann and Pawlak (\citeyear{Bissantz10}).


%

%
%
\section{The Zernike orthogonal basis and image reconstruction}\label
{sec:zernike}
Zernike functions, introduced as an orthogonal and rotationally
invariant basis of polynomials on the disc in Zernike (\citeyear{Zernike34}), and their
corresponding moments have been used extensively in image analysis and
pattern recognition; see Bailey and Srinath (\citeyear{Bailey96}), Khotanzad and Hong
(\citeyear{Khotanzad90}) and Mukundan and Ramakrishnan (\citeyear{Mukundan98}). The Zernike basis has also
been employed as an important tool for the statistical inference
concerning the inverse problem of positron emission tomography
[cf. Jones and Silverman (\citeyear{Jones89}) and Johnstone and Silverman (\citeyear{Johnstone90})] and
PSF estimation in fluorescence microscopy [cf. Dieterlen et al.
(\citeyear{Dieterlen04});
Dieterlen et al. (\citeyear{Dieterlen08})].
%

\subsection{Zernike polynomials}
In the following we identify two-dimensional space~$\mathbb{R}^2$ with the
complex plane $\mathbb{C}$ via $(x,y) \mapsto x + iy$, where $i$ is the
imaginary unit. In particular, $e^{i \beta}$ is the unit vector $(\cos
\beta, \sin\beta)$ at angle $\beta$ to the $x$ axis.

Now, the (complex) Zernike orthogonal polynomials are given by
$ V_{pq}(x,y) = R_{pq}(\rho)   e^{iq\theta}$, $(x,y) \in D$,
where $\rho= \sqrt{x^2+y^2}$, $\theta= \arctan(y/x)$ and $R_{pq}(\rho)$ is the radial Zernike polynomial
given explicitly by
\[
R_{pq}(\rho) = \sum_{l=0}^{(p - |q|)/2} \frac{(-1)^l (p-l)! \rho
^{p-2l}}{l! ((p+|q|)/2 -l )! ((p-|q|)/2 -l )!}.
\]
The indices $(p,q)$ have to satisfy
$p \geq0$, $|q| \leq p$, and $p-|q|$ has to be even.
We will call such pairs $(p,q)$ admissible. The Zernike polynomials
satisfy the following orthogonality relation over the unit disc $D$:
\[
\iint_{D} V_{pq}(x,y) V_{p'q'}^*(x,y)  \,dx  \,dy = \pi/(p+1) \delta_{p
p'} \delta_{q q'},
\]
where $^*$ denotes complex conjugation and $\delta_{pp'}$ is the
Kronecker delta. This implies that
\begin{equation}\label{eq:normzernike}
\| V_{pq} \|^2 = \pi/(p+1) = n_p,
\end{equation}
where $\| \cdot\|$ is the norm on $L_2(D)$. In Bhatia and Wolf (\citeyear{Bhatia54}),
the Zernike polynomials are characterized by a certain uniqueness
property, among others, invariant polynomials defined on $D$.
%
\subsection{Function approximation}
Since the family $\{V_{pq}(x,y)\}$ for admissible $(p,q)$ forms a
complete and orthogonal system in $L_2(D)$, we can expand a function $f
\in L_2(D)$ into a series of the Zernike polynomials, that is,
\begin{equation}\label{eq:zernikeexpansion}
f(x,y)= \sum_{p=0}^{\infty} \sum_{q=-p}^p n_p^{-1}  A_{pq}(f) V_{pq}(x,y),
\end{equation}
where here and throughout the paper the summation is taken over
admissible pairs $(p,q)$. Thus, the Fourier coefficients $\{A_{pq}(f)\}
$ (often referred to as the Zernike moments)
uniquely characterize the image function $f$. The norming factor
$n_p^{-1}$ arises due to (\ref{eq:normzernike}), and the Zernike moment
$A_{pq}(f)$ is defined by
\[
A_{pq}(f) = \iint_{D} f(x,y)  V_{pq}^*(x,y)   \,dx \,dy.
\]
Owing to Parseval's formula, we have that for $f \in L_2(D)$
\begin{equation}\label{eq:normexpansion}
\|f\|^2 = \sum_{p = 0}^{\infty} \sum_{q=-p}^p n_p^{-1} |A_{pq}(f)|^2.
\end{equation}
Let us introduce the notation
$ \tilde f(\rho,\theta) = f(\rho\cos\theta, \rho\sin\theta)$
for a function $f \in L_2(D)$. Then by using polar coordinates we obtain
\begin{eqnarray}\label{eq:thezernikecoeff}
A_{pq}(f) &=& 2 \pi\int_0^1 c_q(\rho,f)  R_{pq}(\rho)  \rho
\,d\rho,
\nonumber
\\[-8pt]
\\[-8pt]
\nonumber
c_q(\rho,f)&=& \frac{1}{2 \pi} \int_0^{2 \pi} \tilde f(\rho
,\theta
) e^{-iq \theta} \, d\theta.
\end{eqnarray}
%
%
\subsection{Image reconstruction}
We assume that the data are observed on a symmetric square grid
of edge width $\Delta$, that is, $x_i - x_{i-1} = y_i - y_{i-1} =
\Delta
$ and $x_i = -x_{m-i+1}$, $y_i = -y_{m-i+1}$, so that $(x_i,y_j)$ is
the center of the pixel
$\Pi_{ij}= [x_i-\frac{\Delta}{2}, x_i+\frac{\Delta}{2}]\times
[y_j-\frac
{\Delta}{2},y_j+\frac{\Delta}{2}]$. Note
that $m$ corresponds to $2/\Delta$.
For $f \in L_{2}(D)$ we shall assume the following observational model:
\begin{equation}
\label{eq:regmodel} Z_{i,j} = f(x_i,y_j) + \epsilon_{i,j},\qquad
(x_i,y_j) \in D, 1 \leq i,j \leq m,
\end{equation}
where the noise process $\{\epsilon_{i,j}\} $ is an i.i.d. random
sequence with zero mean, finite variance $E
\epsilon_{i,j}^2 = \sigma^2$ and finite fourth moment, so that
$Z_{i,j}$ is the datum associated with pixel $\Pi_{ij}$. Note that along
the boundary of the disc, some lattice squares are included (if their
center is in $D$) and some are excluded. When reconstructing $f$, this
gives rise to an additional error, called geometric error in Pawlak and
Liao (\citeyear{Pawlak02}). This error can be quantified by using the celebrated
problem in analytic number theory referred to as lattice points of the circle.
In applications, the datum $Z_{i,j}$ might also correspond to the
average of $f$ over the pixel $\Pi_{ij}$ rather than its value at the
center, in such cases we assume negligible variation of $f$ over $\Pi_{ij}$.
In the following we need to work with a discretized version of the
Zernike moments. Consider weights $w_{pq}(x_i,y_j)$ of the form
\begin{equation}\label{eq:weights1}
\quad w_{pq}(x_i,y_j) = \iint_{\Pi_{ij}} V_{pq}^*(x,y)  \,dx \,dy  \quad \mbox{or} \quad   w_{pq}(x_i,y_j) = \Delta^2 V_{p q}^*(x_i,y_j).
\end{equation}
%
%
Using either version in (\ref{eq:weights1}), we estimate the Zernike
moment $A_{pq}(f)$ by
\begin{equation}\label{eq:estizernike}
\hat A_{pq} = \sum_{(x_i,y_j) \in D} w_{pq}(x_i,y_j) Z_{i,j}.
\end{equation}
For efficient methods for computing the Zernike moments $\hat A_{pq}$,
see, for example, Amayeh et al. (\citeyear{Amayeh05}). Instead of the uniform weights,
one could also use a more sophisticated quadrature rule, particularly
if sharp features of $f$ are expected.
%
%

\section{Reflection estimation}\label{sec:reflecest}
First we investigate the effect that reflecting an image function $f$
has on its Zernike moments. Suppose that $f$ is reflected at a line
along the direction $e^{i \beta}$, $\beta\in[0,\pi)$, and denote the
reflected function by $\tau_\beta f$. Then one easily shows that
$\widetilde{(\tau_\beta f)}(\rho, \theta) = \tilde f(\rho, 2\beta-
\theta)$ and, consequently, using (\ref{eq:thezernikecoeff}),
\begin{equation}\label{eq:coefreflect}
A_{pq}(\tau_\beta f) = e^{-2 i q \beta} A_{p,-q}(f).
\end{equation}
Consider the following assumption.
\begin{assum}\label{assum:uniqueaxis}
Suppose that $f \in L_2(D)$ is invariant under some unique reflection
$\tau_{\beta^*}$.
\end{assum}

Indeed, the composition of two reflections along lines $e^{i \alpha_1}$
and $e^{i \alpha_2}$ is a rotation with angle $2(\alpha_2 - \alpha_1)$.
Thus, $f$ is invariant under a unique reflection if and only if~$f$ is
invariant under some reflection and if $f$ is not invariant under any
rotation.
%
\subsection{Contrast functions}
Our method for estimating $\beta^*$ is based on the expansion (\ref
{eq:normexpansion}) and the invariance property of Zernike moments
expressed by the formula in (\ref{eq:coefreflect}).
We set
\begin{equation}\label{eq:seriesltwo}
M(\beta,f) = \|f - \tau_\beta f\|^2 = \sum_{p=0}^\infty n_p^{-1}
\sum
_{q=-p}^p |A_{pq}(f) - e^{-2 i q \beta} A_{p,-q}(f)|^2.
\end{equation}
Evidently, under Assumption \ref{assum:uniqueaxis} the angle $\beta^*$
is the unique zero of the function $M(\beta,f)$.
Writing $A_{pq}(f) = |A_{pq}(f)| e^{i r_{pq}(f)}$ and noting that
$A_{pq}(f) = A_{p,-q}(f)^*$, we calculate
\begin{equation}\label{eq:characreflec}
M(\beta,f)  = \sum_{p=0}^\infty n_p^{-1} \sum_{q=0}^p 4
|A_{pq}(f)|^2 \bigl(1 - \cos\bigl(2 r_{pq}(f) + 2 q \beta\bigr) \bigr),
\end{equation}
where the sums are taken over admissible pairs $(p,q)$. Therefore,
$\beta^*$ is also uniquely characterized by the condition $\cos(2
r_{pq}(f) + 2 q \beta^*) = 1
$ or by requiring
\begin{equation}\label{eq:estequality}
r_{pq}(f) \in q \beta^* + \pi\mathbb{Z}
\end{equation}
for all $p,q$ with $A_{pq}(f) \not=0$.

Thus, a natural way to estimate $\beta^*$ is to first estimate a
truncated version of the series defining $M(\beta,f)$, and then define
an estimate of $\beta^*$ as the minimizer of this estimated contrast
function. We first show that suitably truncated versions of $M(\beta
,f)$ still uniquely determine $\beta^*$. For a fixed $N$ set
\[
M_N(\beta,f) = \sum_{p=0}^N n_p^{-1} \sum_{q=-p}^p |A_{pq}(f) -
e^{-2 i
q \beta} A_{p,-q}(f)|^2.
\]
This is the truncated counterpart of the series in (\ref
{eq:seriesltwo}). Evidently, $M_N (\beta^*,f)=0$ for all $N$ under
Assumption \ref{assum:uniqueaxis}.
We shall call $M(\beta,f)$ and $M_N(\beta,f)$ as well as their
empirical version below \textit{contrast functions}.
\begin{theorem}\label{prop:uniquebeta}
Suppose that $f$ satisfies Assumption \ref{assum:uniqueaxis}. Then for
sufficiently large $N = N(f)$, $\beta^*$ is the \textit{unique zero} of
the truncated contrast functions $M_N(\beta,f)$.
\end{theorem}
The proof of Theorem \ref{prop:uniquebeta}, given in the supplementary material in
Bissantz, Holzmann and Pawlak (\citeyear{Bissantz10}), reveals that in order to uniquely
determine the direction of the reflection axis $e^{i \beta^*}$ as the
zero of the function $M_N(\beta,f)$ one has to choose~$N$ so large such
that the sum defining $M_N(\beta,f)$ contains nonzero $A_{pq}$'s for
which the greatest common divisor (gcd) of the $q$'s is $1$. Thus, we
can choose $N$ as the smallest value such that $A_{p_1q_1}(f)\neq0,$
$\ldots,A_{p_rq_r}(f)\neq0,$ for $p_i \leq N, i=1,\ldots,r$, with
$\operatorname{gcd}(q_1,\ldots, q_r) =1.$

In practice, $M_N(\beta,f)$ and hence an appropriate value for $N$
still has to be estimated. One could test sufficiently many moments to
be nonzero, however, we prefer to choose $N$ for appropriate estimation
of $f$ in the resulting truncated Zernike series estimate; see Section \ref{sec:simu}.
Apart from its theoretical value, Theorem \ref{prop:uniquebeta}
implies that even in dim images occurring, for example, in PSF
estimation in Section \ref{sec:simu}, where only few Zernike moments
may be properly estimated, it is still possible to identify and
estimate the symmetry axis.
%
\subsection{Estimation}
For estimation purposes, we first estimate the contrast functions
$M_N(\beta, f)$ by
\[
\hat M_{N}(\beta)  =  \sum_{p=0}^N n_p^{-1} \sum_{q=-p}^p |\hat
A_{pq} - e^{-2 i q \beta} \hat A_{p,-q}|^2,
\]
where we write $\hat A_{pq} = |\hat A_{pq}| e^{i \hat r_{pq}}$.
Then we define the estimator of $\beta^*$ as
\[
\hat\beta_{\Delta,N} = \operatorname{arg} \min_{\beta\in[0, \pi)} \hat
M_N(\beta).
\]
The estimate $\hat\beta_{\Delta,N}$ depends on the grid size $\Delta$
and, more importantly, on the truncation parameter $N$. Note that
although $M_N(\beta^*)=0$, $\hat M_N(\hat\beta_{\Delta,N})$ will be
positive a.s. due to noise.
\begin{rem}
The estimated contrast function $\hat M_N(\beta)$ is simply the squared
$L_2$ distance between the Zernike estimate given in polar coordinates by
\[
\widetilde{{\hat f}} (\rho, \theta) = \sum_{(p,q)}^N n_p^{-1} \hat
A_{p,q} \tilde V_{p,q}(\rho, \theta)
\]
and its reflected version $\tau_\beta\hat f$. Note that this is
achieved by a special property of the Zernike polynomials, namely, the
set of Zernike polynomials used in the estimate~$\hat f$ remains
invariant under reflection.
As suggested by a referee, an estimate similar to $\hat\beta_{\Delta
,N}$ would be obtained by estimating the coefficients $A_{p,q}$ in
\begin{equation}\label{eq:linearmodel}
\tilde f (\rho, \theta) - \tau_\beta  \tilde f (\rho, \theta) =
\sum
_{(p,q)}^N n_p^{-1} ( A_{p,q} - A_{p,-q} e^{-2 \beta i q})
\tilde V_{p,q}(\rho, \theta)
\end{equation}
by least squares for each fixed $\beta$, and then choosing the $\beta$
with minimal RSS. While our approach is somewhat simpler since we
estimate the $A_{p,q}$ \textit{before} imposing symmetry (and thus
independently of $\beta$), this approach could potentially be placed
into a likelihood or Bayesian framework as in Pankajakshan et al. (\citeyear{Pankajakshan08}).
\end{rem}

The next result states uniform convergence in probability of the
estimated contrast function $\hat M_N(\beta)$ to $M_N(\beta)$. This is
also used in order to obtain the consistency of $\hat\beta_{\Delta,N}$
for $\beta^*$.
\begin{theorem}\label{prop:uniformconvcont}
For each fixed $N$, as $\Delta\to0$,
\begin{equation}\label{eq:uniformconv}
\sup_{\beta\in[0,\pi)} |\hat M_N(\beta) - M_N(\beta,f)| \to0\qquad
 (P),
\end{equation}
where $(P)$ denotes convergence in probability.
\end{theorem}

Note that in Theorem \ref{prop:uniformconvcont}, $f$ need not be
reflection invariant.
The next theorem gives the consistency of $\hat\beta_{\Delta,N}$ as
$\Delta\to0$ as well as its parametric $\Delta$-rate of convergence.
In all the results that follow we choose the truncation parameter $N$
according to the prescription established in Theorem \ref{prop:uniquebeta}, that is, we
require that $N$ should
be selected in such a way that $\beta^*$ is the unique minimizer of
$M_N(\beta,f).$ We will refer to such a value of $N$ as ``sufficiently large.''
\begin{theorem}\label{lem:consistency}
Suppose that $f \in L_2(D)$ is a function of bounded variation and
satisfies Assumption~\ref{assum:uniqueaxis}. Then for sufficiently
large (but fixed) $N = N(f)$, we have that, as $\Delta\to0$,
\begin{equation}\label{ratedelta}
|\hat\beta_{\Delta,N} - \beta^*| = O_P(\Delta).
\end{equation}
\end{theorem}

Next we establish asymptotic normality for the estimate $\hat\beta
_{\Delta,N}$. In order for the bias term of the estimated Zernike
coefficient to be negligible, we require that the image function $f$ is
Lipschitz continuous.
\begin{theorem}\label{the:asympnorm}
Suppose that $f$ is Lipschitz continuous and satisfies Assumption~\ref
{assum:uniqueaxis}. Then for sufficiently large (but fixed) $N = N(f)$,
we have that, as $\Delta\to0$,
\begin{equation}\label{eq:asympnorm}
\Delta^{-1}  (\hat\beta_{\Delta,N} - \beta^* )
\stackrel
{\mathcal{L}}{\to}  N\biggl(0, \frac{8 \sigma^2}{M_N''(\beta
^*,f)}\biggr),
\end{equation}
where
\begin{equation}\label{eq:thelimitvariance}
M_N''(\beta^*, f) = \sum_{p=0}^N n_p^{-1} \sum_{q=0}^p 16
|A_{pq}(f)|^2 q^2 .
\end{equation}
\end{theorem}

Theorem \ref{the:asympnorm} can be used to construct an asymptotic
confidence interval for $\beta^*$. To this end, we need an estimate of
the asymptotic variance in the normal limit (\ref{eq:asympnorm}). We
may estimate $M_N''(\beta^*, f)$ directly by using (\ref
{eq:thelimitvariance}) simply by replacing $A_{pq}(f)$ by $\hat
A_{pq}$. However, this may result in underestimation of the asymptotic
variance, and therefore plugging $\hat\beta_{\Delta,N}$ into the
second derivative of $\hat M_N(\beta$),
\[
\hat M_N''(\beta) = \sum_{p=0}^N n_p^{-1} \sum_{q=0}^p 16 |\hat
A_{pq}|^2 q^2 \cos(2 \hat r_{pq} + 2 q \beta),
\]
should generally be preferred. Call either estimate $\hat M_N''$.
Further, we need to estimate the error variance $\sigma^2$. To this
end, one could use the residuals from the fitted truncated Zernike
series. We prefer to use a difference estimate of the form %
\begin{equation}\label{eq:varest}
\hat\sigma^2 = \frac{1}{C(\Delta)} \sum_{(x_i,y_j) \in D} \frac{1}{4}
\bigl((Z_{i,j}- Z_{i+1,j} )^2 + (Z_{i,j}- Z_{i,j+1} )^2
\bigr),
\end{equation}
which does not rely on the same underlying regression estimate. Here
the sum is taken over all $(x_i,y_j) \in D$ where $(x_{i+1},y_j) \in D$
and $(x_i,y_{j+1}) \in D$, and $C(\Delta)$ is the number of terms in
this restricted sum. One can show that if $f$ is Lipschitz continuous,
then $ \hat\sigma^2 - \sigma^2 = O_P(\Delta)$. For detailed
information on difference-based estimators in higher dimensions see
Munk et al. (\citeyear{Munk05}).

Using these estimates, we obtain the following confidence interval with
nominal level $\alpha$ for $\beta^*$:
\begin{equation}\label{eq:cb}
\biggl[ \hat\beta_{\Delta,N} - u_{1-\alpha} \cdot  \frac{2 \sqrt{2}
\hat\sigma\Delta}{ (\hat{M}_N'')^{1/2}},   \hat\beta_{\Delta,N} +
u_{1-\alpha} \cdot \frac{2 \sqrt{2} \hat\sigma\Delta
}{(\hat{M}_N'')^{1/2}}\biggr],
\end{equation}
where $u_{1-\alpha}$ is the $1-\alpha$-quantile of the standard normal
distribution.
\begin{rem}
If in Theorem \ref{the:asympnorm} we only assume that $f \in L_2(D)$ is
a function of bounded variation, then the bias is also of order $\Delta
$, and we get an asymptotic offset (i.e., a limiting normal law with
nonzero mean) in (\ref{eq:asympnorm}).
\end{rem}
\begin{rem}
If the image $f$ is not reflection invariant, the estimator $\hat\beta
_{\Delta,N}$ may still converge to a certain parameter value $\beta
^\diamond$, which is determined by minimizing the $L_2$-distance $\|f -
\tau_\beta f\|^2$. Then $(f + \tau_{\beta^\diamond} f)/2$ is the best
reflection-symmetric approximation (in the $L_2$ sense) to the original
image $f$. However, since $\beta^\diamond$ is no longer a zero of the
contrast function $M(\beta, f)$, Theorem \ref{prop:uniquebeta} does not
hold, and in order to achieve consistent estimation theoretically, one
requires that $N \to\infty$.
\end{rem}
\begin{rem}\label{rem:notunique}
Suppose that $f$ is reflection invariant but is also invariant under
some discrete rotation group. Then there will be a minimal angle
$\alpha= 2 \pi/d$ for some $d \in\mathbb{N}$, under rotation of
which $f$ is
invariant. If we use the estimator $\hat\beta_{\Delta,N}$ in such a
situation, then a unique reflection axis will be between $0$ and
$\alpha
$, and\vspace*{1pt} one should use the minimizer of $\hat M_N(\beta)$ in the
interval $[0, \alpha)$ rather than in $[0, \pi)$.
\end{rem}

\section{Finite sample performance}\label{sec:finitesample}

\subsection{Target functions and the shape of their contrast functions}

\begin{figure}
\includegraphics{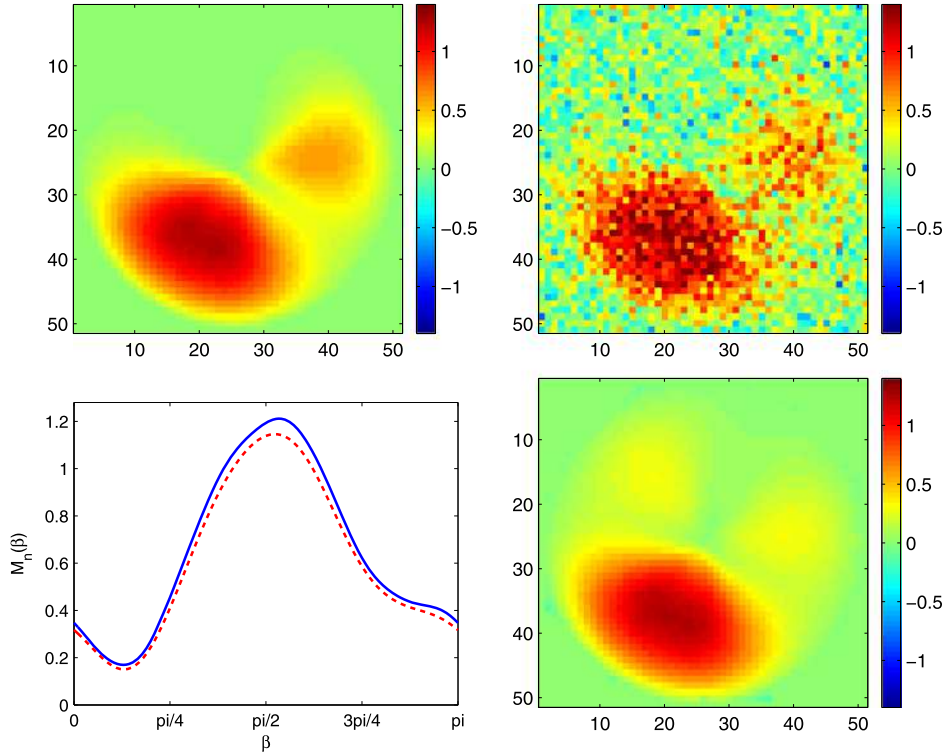}
\caption{The function $f_3$ (which is \textit{not}
reflection symmetric) without noise, with Gaussian noise, and
$M_7(\beta
)$ (full curve), $\hat M_7(\beta)$ (dashed curve) and a symmetrized
version of $f_3$.
Parameters are $n=25$ and signal-to-noise-ratio${}=5$.}\label{fig:f3}
\end{figure}

\begin{figure}
\includegraphics{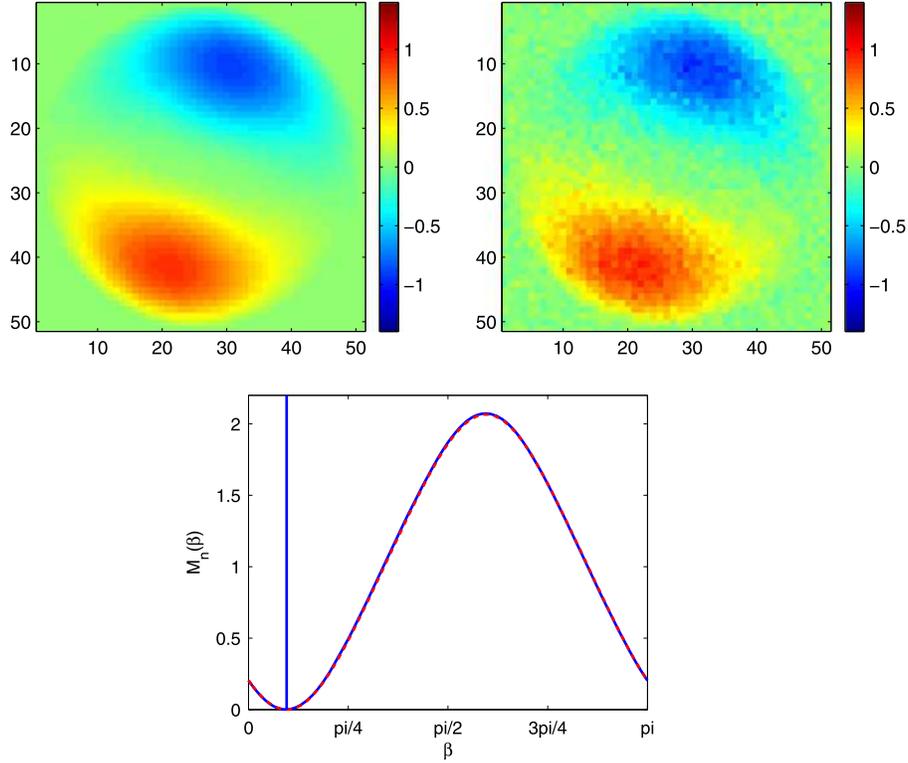}
\vspace*{-6pt}
\caption{The reflection symmetric function $f_1$ without
noise, with Gaussian noise, and $M_7(\beta)$ (full curve) and $\hat
M_7(\beta)$ (dashed curve).
Parameters are $n=25$ and signal-to-noise-ratio${}=16.7$. The vertical
line indicates the direction of reflection symmetry in the true image.}\label{fig:f1b}
\vspace*{-6pt}
\end{figure}

In this section we discuss the results of a simulation study of the
proposed estimation method
for the angle $\beta$ of reflectional symmetry. We performed
simulations with three target functions, which are given in
polar coordinates by
\begin{eqnarray*}
 f_1(\rho,\theta) & = & c_1\cdot x\cdot(1-\rho)\cdot\bigl(\sin
\bigl(y+\sqrt{x^2+y^4}\bigr)+\sin\bigl(-y+\sqrt{x^2+y^4}\bigr)\bigr),\\
f_2(\rho,\theta) & = & c_2 \cdot\rho\cdot(1-\rho)\cdot
\bigl(e^{\cos
(\theta)/0.02}+e^{\cos(\theta+0.6)/0.02}\\
&&\hspace*{51pt}\hspace*{22pt}{}+e^{\cos(\theta-0.3+\pi
)/0.02}+e^{\cos(\theta+0.9+\pi)/0.02}\bigr),\\
f_3(\rho,\theta) & = & c_3\cdot\rho\cdot(1-\rho)\cdot
\bigl(e^{\cos
(\theta)/0.2}+e^{\cos(\theta+0.9)/0.2}+0.6\cdot e^{\cos(\theta
-1.7)/0.2}\bigr),
\end{eqnarray*}%
%
\begin{figure}[b]
\vspace*{-6pt}
\includegraphics{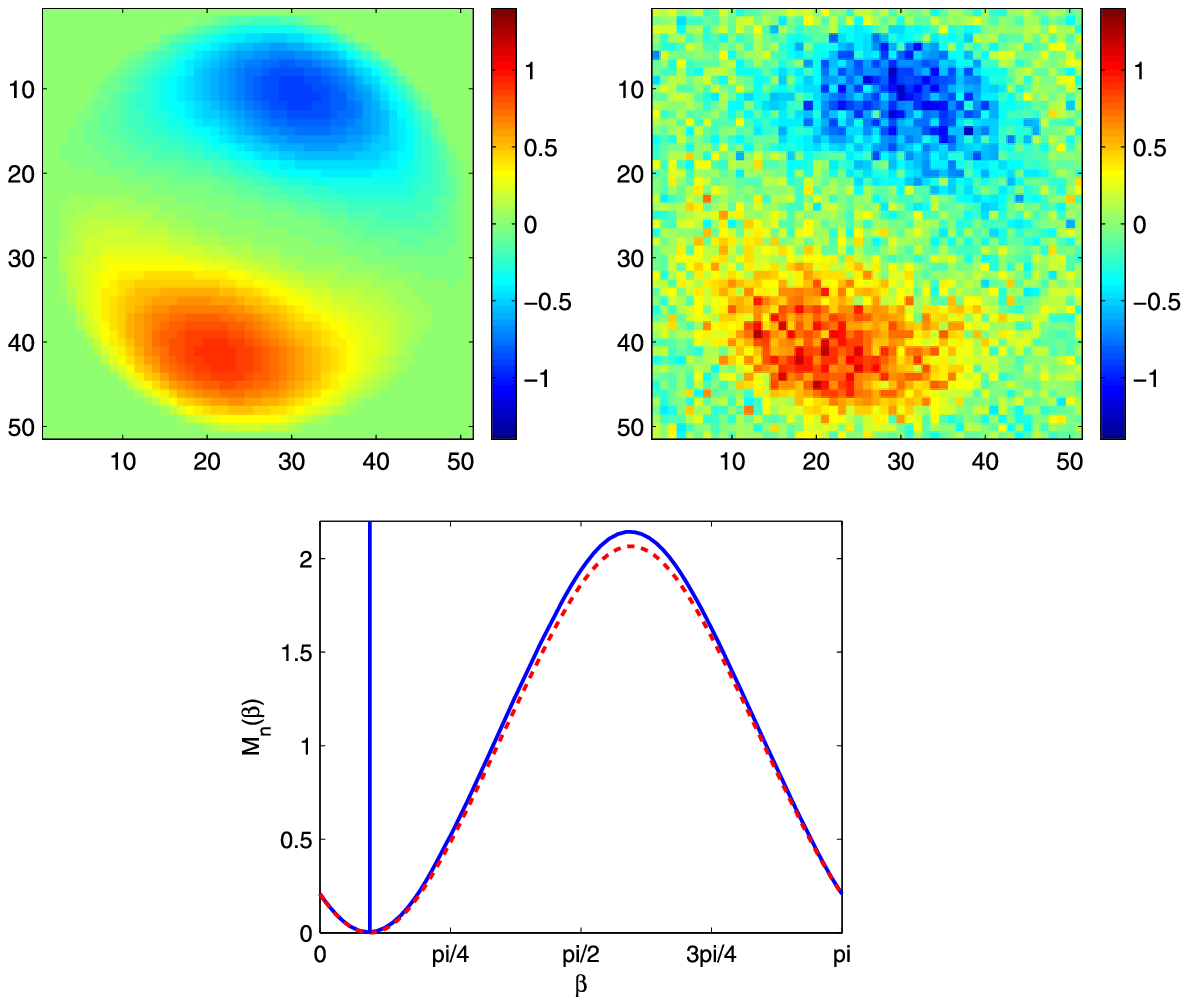}
\caption{Reflection symmetric function $f_1$ without
noise, with Gaussian noise, and $M_7(\beta)$ (full curve) and $\hat
M_7(\beta)$ (dashed curve).
Parameters are $n=25$ and signal-to-noise-ratio${}=5$. The vertical line
indicates the direction of reflection symmetry in the true image.}\label{fig:f1}
\end{figure}%
%
\begin{figure}
\includegraphics{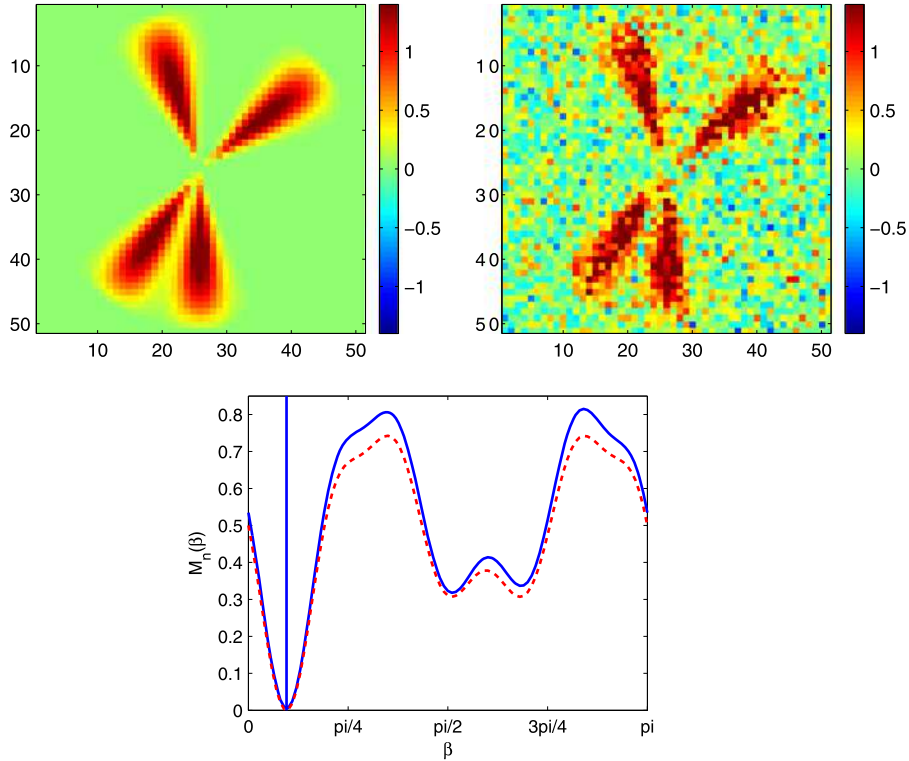}
\caption{Reflection symmetric function $f_2$ without
noise, with Gaussian noise, and $M_7(\beta)$ (full curve) and $\hat
M_7(\beta)$ (dashed curve).
Parameters are $n=25$ and signal-to-noise-ratio${}=5$. The vertical line
indicates the direction of reflection symmetry in the true image.}\label{fig:f2}
\vspace*{-6pt}
\end{figure}%
where $x=\rho\cos(\theta)$, $y=\rho\sin(\theta)$, and
$c_1,c_2,c_3$ are
normalization constants such that the squared functions all integrate
to one on the unit disc.
Figures \ref{fig:f1}--\ref{fig:f3} show the target functions without
noise and with Gaussian noise, where the signal-to-noise ratio, defined
as the ratio between the peak values of the respective target function
$f_1,f_2,f_3$ and the standard deviation of the noise $\sigma$, is $5$.
Figure \ref{fig:f1b} again shows $f_1$ but with a signal-to-noise ratio
of $16.7$.
Note that the
functions $f_1$ and $f_2$ are reflection symmetric, whereas $f_3$ is
not. Moreover, in all cases we have used regularization parameters $N$
chosen according to the selection rule described in Bissantz, Holzmann
and Pawlak (\citeyear{Bissantz09}) (a stochastic analogue of the numerical discrepancy
principle for parameter selection in inverse problems). The fact that
$f_1$ and $f_2$, in contrast to $f_3$, are reflection symmetric is
clearly expressed in the shape of the associated contrast functions.
Indeed, $\hat M_7(\beta)$ is far above zero for $f_3$, in contrast to
the case of $f_1$ and $f_2$, where $\hat M_7(\beta)$ reaches a minimum
close to zero for noisy data. However, we note that even for $f_3$
there still exists a well-defined minimum
of the contrast function $\hat M_7(\beta)$. The right panel in Figure
\ref{fig:f3} shows a reflection symmetric version of $f_3$, which has
been generated by adding a version of $f_3$ mirrored w.r.t. the axis
given by the direction of the minimum of $\hat M_7(\beta)$.

\subsection{Simulated distributions of estimated directions $\hat\beta$}

\begin{sidewaysfigure}
\includegraphics{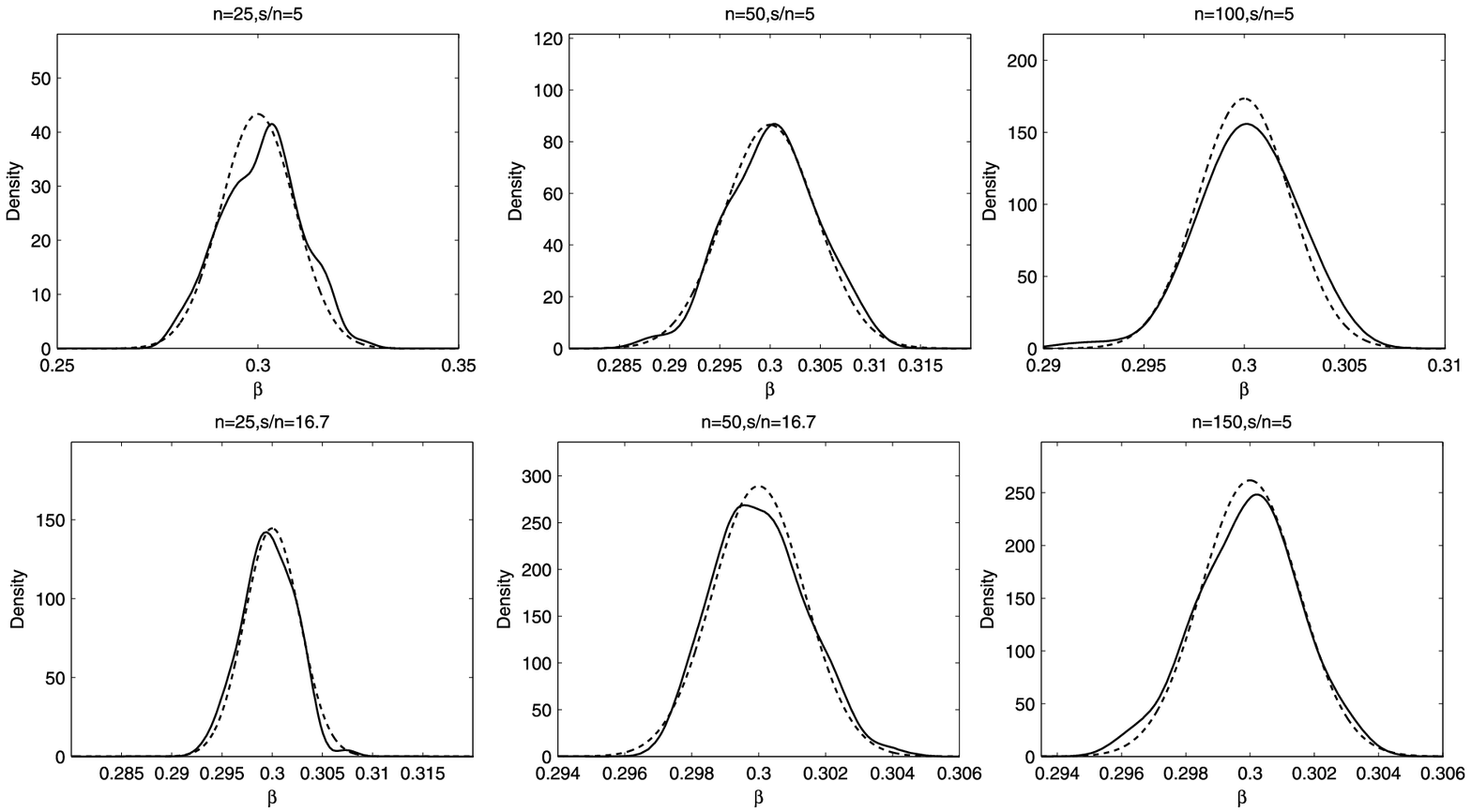}
\caption{Simulated (solid curve) and asymptotic
(dashed curve) distributions of $\hat\beta_{\Delta,N}$ for $f_1$. The
variance of the asymptotic distributions is given as $\frac{8\sigma
^2\Delta^2}{M''_N(\beta^{\ast})}$ (cf. Theorem \protect\ref{lem:consistency}).
The parameter $N$
was chosen as $7,8,12$ (first row, left to right) and $8,12,12$ (second
row, left to right).}\label{fig:betasimu1}
\end{sidewaysfigure}

\begin{figure}[t]
\includegraphics{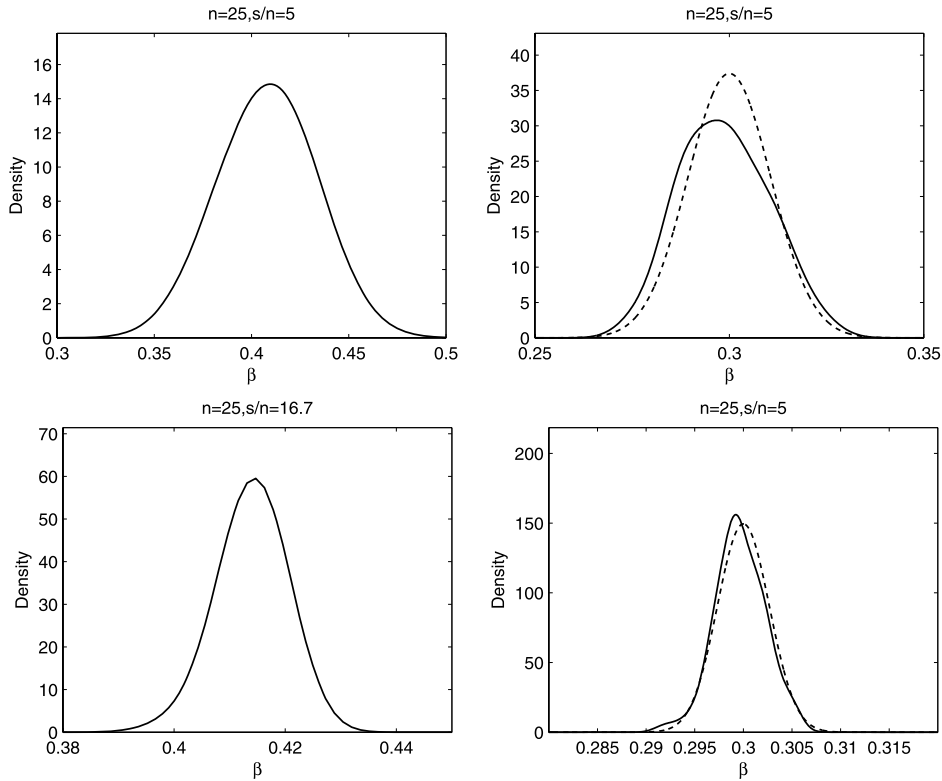}
\caption{Simulated (solid curve) and asymptotic
(dashed curve) distributions of $\hat\beta_{\Delta,N}$ for $f_2$ (left
panels) and $f_3$ (right panels). The variance of the asymptotic
distributions is given as $\frac{8\sigma^2\Delta^2}{M''_N(\beta
^{\ast
})}$ (cf. Theorem \protect\ref{lem:consistency}).
The parameter $N$ was chosen as $7$, except for
the lower left where it was chosen as $8$.}\label{fig:betasimu2}
\end{figure}

In the second part we have simulated the distribution of $\hat\beta$,
determined as the minimum of
$\hat M_N(\beta)$, for a range of values for the parameters $n$ and the
signal-to-noise ratio $s/n$. Figures~\ref{fig:betasimu1} and~\ref
{fig:betasimu2} show density plots of the simulated distributions
together with normal limits. For the reflection symmetric functions
$f_1$ and $f_2$ we compare the simulated distributions to their
asymptotic counterparts according to (\ref{eq:asympnorm}). Even for
images of moderate size such as the unit circle in the square image
with edge length $(2m+1)=51$ pixels, the simulated distributions are
already close to their asymptotic limit.

\begin{figure}
\includegraphics{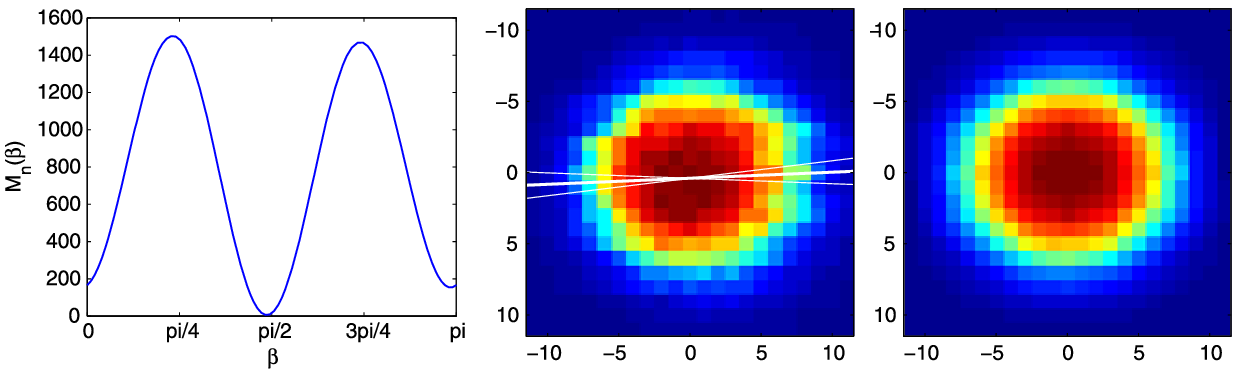}
\caption{Contrast function of image bead (left), image bead with superposed
estimated reflection axis and the associated asymptotic confidence
interval with nominal level $95\%$ (middle), and bead after averaging
along two estimated axes of reflectional symmetry (right). Bead was
acquired during two observation runs of HeLa cervix carcinoma cells
with a Leica TCS laser scanning fluorescence microscope. Here, we used $N=4$.}\label{fig:bead2}
\end{figure}

\section{Calibrating the PSF in confocal microscopy}\label{sec:simu}

\subsection{Assessing reflectional symmetry of the PSF}
\label{sec:estaxis}

In this section we use the contrast function to estimate the axes of
reflection symmetry in an image of the point-spread function
in confocal fluorescence microscopic imaging. Here one observes count
data representing observed pixel-integrated image intensities on a
two-dimensional (or three-dimensional) equidistant grid of pixels. We
consider the two-dimensional case,
where the observations are
$Z_{i,j} = (K\gamma)(x_i,y_j) + \varepsilon_{ij}$,
with
\begin{equation}\label{eq:conv}
(K \gamma) (x,y) = k \ast\gamma(x,y) = \int_{\mathbb{R}^2}
k(x-t_1, y-t_2)
\gamma(t_1,t_2)  \,dt_1 \,dt_2,
\end{equation}
and where ``$\ast$'' represents the convolution of the ``true'' image
$\gamma\in L^2$ with the so-called point-spread-function (PSF) $k\in
L^2$ of the microscope. The standard model for the distribution of the
photon count data $Z_{i,j}$ is that $Z_{i,j}$ is Poisson with the mean
$ (K\gamma)(x_i,y_j)$, all independent.

The PSF represents the image of a point-source observed by the
microscope and describes the blurring effect of the imaging process. As
discussed in the introduction, the PSF is typically estimated by
observing a point-like object (called bead) of known form. Figure \ref
{fig:bead2} (right) shows the image of a bead under a Leica TCS
confocal laser scanning microscope.\ The (observed) empirical PSF is
typically no longer rotationally invariant, but it often remains
reflection symmetric under two (unknown) orthogonal axes, even if, for
example, the detector plane was not in perfect agreement with the focal
plane of the microscope; cf. Lehr, Sibarita and Chassery (\citeyear{lehr98}) and Pankajakshan et al. (\citeyear{Pankajakshan08}).

In Bissantz, Holzmann and Pawlak (\citeyear{Bissantz09}) we applied tests both for
rotational invariance and for invariance under a rotation by $\pi$
(which is an immediate consequence of reflection symmetry w.r.t. two
orthogonal axes) to the observed PSF in the image bead (Figure \ref
{fig:bead2}). It turned out that rotational invariance could be
rejected at a 5\% level, but invariance under a rotation by $\pi$ was
not rejected.

For a deeper investigation, we now apply our methodology to estimate
the (orthogonal) axes of reflection symmetry.
The data from fluorescence microscopic imaging in general is
distributed (approximately) according to a Poisson distribution with
expectation given by the respective image intensity.\ Hence, the noise
is not homoscedastic as required by model (\ref{eq:regmodel}). As
suggested by a referee, we use the (variance stabilizing) Anscombe
transform [Anscombe (\citeyear{Anscombe48})]. Note that reflection symmetry is preserved
in this process. Further, following Remark \ref{rem:notunique}, we
restrict the range of $\beta$ to $[\pi/4, 3\pi/4]$, which yields an
estimated angle and associated $95\%$ confidence interval of $\hat
\beta
=1.54\pm0.07$. The truncation parameter was selected as $N=4$ by the
method described in Bissantz, Holzmann and Pawlak (\citeyear{Bissantz09}).
Using the untransformed data and ignoring heteroscedasticity yields
quite similar results ($\hat\beta=1.53\pm0.08$), thus,
heteroscedasticity appears to be a minor problem in this context.

Figure \ref{fig:bead2} (left and middle) shows the contrast function
(of the untransformed data with $N=4$) and the image with superposed
estimated reflection axis ($\hat\beta\approx1.53$).
The optical axis appears to be rather close to the coordinate axis of
the image. In particular, the coordinate axis is covered by the
associated $95\%$-nominal level confidence interval for $\hat\beta$
[cf. (\ref{eq:cb})]. This indicates that the reason for a PSF which is
not rotationally invariant appears to be some (slight) misalignment of
the optical system or anisotropy of the
immersion medium used for object preparation rather than some random
deviation from sphericity of the bead used to image the PSF. In Figure
\ref{fig:bead2} (right), we plot the PSF after averaging along two
estimated axes of reflectional symmetry.
%

\subsection{Performance of symmetrized PSF estimates for image reconstruction}

In this section we discuss the results from an extensive simulation
study in which we investigate the
potential benefit of incorporating symmetry information into PSF estimates.

We shall compare the performance of several models for the PSF for
subsequent image reconstruction in a two-step simulation procedure
which mimics the observational process in confocal microscopy. In the
first step, we generate an image of a point-like object and use it to
estimate the PSF in the distinct model classes. In the second step,
these estimated PSFs are employed to reconstruct (by deconvolution with
the estimated PSFs) a target image, and the accuracy of the resulting
reconstructions is compared.

Inference on the PSF as required in the first step has to be conducted
from dim images, and hence requires low-dimensional modeling.
A possible approach is to use a parametric model; however, this
involves the risk of misspecification.
As an alternative, one could seek nonparametric estimates for the PSF.
Due to the dimness of the image, nonparametric smoothing algorithms
would require a substantial amount of smoothing. Therefore, the
essential local feature of the PSF, the steep central peak, would be
reduced, and hence its optical transfer function would be distorted.
Thus, as an actual estimate of the PSF for the reconstruction process,
the Zernike series estimates or other smoothed estimates should not be used.
However, we argue that even for a dim image the Zernike estimate with
few Zernike moments can be used for recovering the global feature of
reflectional symmetry. Averaging along the estimated axes then reduces
the noise level in the PSF reconstruction, which improves the
reconstruction in step 2.

Specifically, the true PSF in the first step in the simulations
consists of a
bivariate Gaussian density function with full width at half maximum
[FWHM] of $250$ nm along
the $y$-axis and
$250/\sqrt{2}$ nm along the $x$-axis, and the bead used to estimate the
PSF is assumed to be $50$ nm in diameter. Moreover, the (true) peak intensity
in the image
of the bead is $\approx$22, which yields a signal-to-noise ratio for the
brightest pixels of $\approx$5.

We use four models in which we estimate the PSF from the available
(Poisson-distributed) observations. First, we use two parametric
models, one correctly specified (i.e., the intensities have the shape
of a Gaussian density with unknown covariance matrix), the other
slightly misspecified with intensity function proportional to
$\exp(\mbox{--}1/2 (q(x,y))^{0.95})$ where $q(x,y) = (x,y) \Sigma^{-1} (x,y)^T$.
Both models are estimated by maximum likelihood. Further, we use two
nonsmoothed nonparametric estimates. The first simply consists of the
observed raw data, for the second we average the raw data along the two
estimated axes of reflectional symmetry, thereby reducing the noise
level by a factor 2.

\begin{figure}

\includegraphics{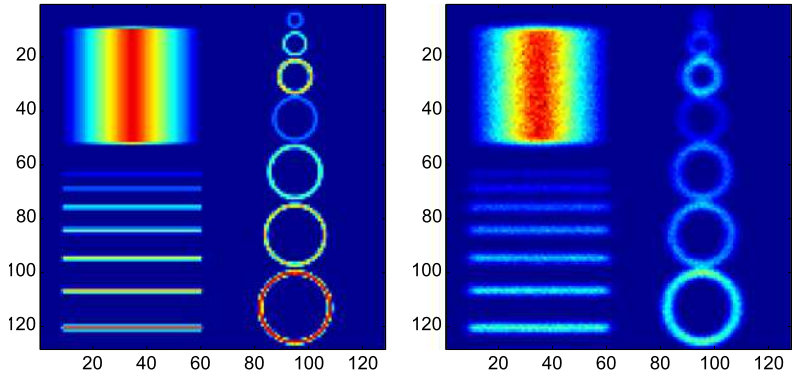}

\caption{Test image used in the simulations of the benefit from using an
estimated axis
of symmetry for the PSF. Left: true image; right: convolved image with
Gaussian noise.}\label{fig:testimage}\vspace*{3pt}
\end{figure}

In the second step we aim to recover the target image plotted in Figure
\ref{fig:testimage} from Poisson-observations with intensities given in
(\ref{eq:conv}), that is, the convolution of the target image and the
true PSF described above.
The target image is of size 8.2 {\textmu}m along the $x$
and $y$-directions and with $128\times128$ pixels along each axis,
that is, the
resolution of a
pixel is $\approx$64 nm. The
signal-to-noise ratio of the brightest pixels is $\approx$20 (and
correspondingly lower for most of the image).
For the image reconstruction by deconvolution, the distinct estimated
PSFs are employed in the same algorithm. We use the
Expectation Maximization method [cf. Shepp and Vardi (\citeyear{Shepp82})], also
called the Richardson--Lucy\vadjust{\goodbreak}
algorithm [cf. Richardson (\citeyear{Richardson72}) and Lucy (\citeyear{Lucy74})],
which is one of the most commonly used algorithms for deconvolution problems
with positivity constraint.
For each estimate of the PSF we record the smallest $L_1$- and
$L_2$-distances attained between any
iterate of the
Richardson--Lucy reconstruction based on the respective PSF and the true
target image in Figure~\ref{fig:testimage}.
%

%
%

%
\begin{table}[b]
\caption{Mean optimal $L_1$- and $L_2$-distance achieved
between reconstructed image and true image, based on
three different estimates of the PSF}\label{tab:PSFsimu}
\begin{tabular*}{\textwidth}{@{\extracolsep{\fill}}lcccc@{}}
\hline
\textbf{Distance} &\textbf{Parametric} & \textbf{Parametric} & \textbf{Nonparametric} &
\textbf{Nonparametric} \\
\textbf{measure} & & \textbf{(misspecified)} & \textbf{without symmetry} & \textbf{with
symmetry} \\
\hline
$L_1$ $(\times 10^5)$ & $1.3$ & $1.9$ & $2.0$ & $1.7$\\
$L_2$ $(\times 10^7)$ & $0.6$ & $1.6$ & $1.5$ & $1.2$\\
\hline
\end{tabular*}
\end{table}

Table \ref{tab:PSFsimu} shows the mean optimal $L_1$- and
$L_2$-distances from
$200$ simulations of the imaging process, that is, subsequent execution
of steps 1
and 2.
It turns out that while the correctly specified parametric model
performs best for recovering the target image, symmetrizing the
nonparametric estimate greatly improves its performance, even beyond
that of the slightly misspecified parametric model.

\section{Conclusions}\label{sec:conclusions}

Detection and estimation of symmetry are fundamental concepts in many
areas of science and technology. In particular, the concept of symmetry
plays an important role in image analysis and pattern recognition.

Symmetry is also relevant in many statistical models.
An important and well-studied example is the symmetric location model
$h(x-\theta),$ where $h(x)=h(-x)$ is an unknown\ symmetric density
function and $\theta\in\mathbb{R}$ is the location parameter. Such models
consisting of a Euclidean parameter as well as a nonparametric
component are called semiparametric, and efficient, that is,
asymptotically optimal estimation procedures in such problems are
important and difficult issues in statistical inference [Bickel et al. (\citeyear{Bickel93})].

In this paper we have discussed how to estimate the angle of the axis
of reflectional symmetry of an image function, and studied its
asymptotic properties. This problem is also of a semiparametric form,
with the angle $\beta$ as the target parameter, and the image function
(that is reflection symmetric with respect to a fixed axis, say, the
$x$-axis) as nonparametric component. Although we showed that the
parametric rate is achievable for estimating the parameter $\beta$, and
also obtained an asymptotic normal law, we did not go into the problem
of semiparametric efficiency and leave this issue for future
research.

We have applied our method to calibrating the point-spread
function (PSF) in confocal microscopy. In particular, we have shown how
reflection symmetry (but no rotational invariance) may arise in the
PSF. Further, we demonstrated that estimating the symmetry axes and
symmetrizing the image of the PSF reduces the noise level in
nonparametric estimates, and can lead to substantial improvement in the
performance in subsequent image reconstruction algorithms.\looseness=1

Future research will be directed toward elucidating symmetry
information and
estimation in more complex microscopic
setups, in particular, in 3D-fluorescence microscopy
[e.g., 4PI-microscopy as in Bewersdorf, Schmidt and Hell (\citeyear{Bewersdorf06})].

\section*{Acknowledgments}
The authors are
indebted to Kathrin Bissantz for her support and advice with the
application to high resolution fluorescence microscopy data. Further,
the authors thank the editor Michael Stein, the associate editor as
well as the reviewers for their helpful comments.
%

\begin{supplement} [id-suppA]
\stitle{Estimating bilateral symmetry: Technical details\\}
\slink[doi]{10.1214/10-AOAS343SUPP} 
\slink[url]{http://lib.stat.cmu.edu/aoas/343/supplement.pdf}
\sdatatype{.pdf}
\sdescription{Here we provide the technical proofs for our results in the paper
``Improving PSF calibration in confocal microscopic imaging---estimating and
exploiting bilateral symmetry.''}
\end{supplement}

\printaddresses

\end{document}